\documentclass[nofootinbib,twocolumn,preprintnumbers]{revtex4-1}
\pdfoutput=1
\usepackage{amsmath,amsthm,amssymb,multirow,psfrag}
\usepackage{epsfig}
\usepackage{color}

\graphicspath{{./Figures/}}

\begin{document}

\def\lsim{\mathrel{\rlap{\lower4pt\hbox{\hskip1pt$\sim$}}
    \raise1pt\hbox{$<$}}}
\def\gsim{\mathrel{\rlap{\lower4pt\hbox{\hskip1pt$\sim$}}
    \raise1pt\hbox{$>$}}}
\newcommand{\vev}[1]{ \left\langle {#1} \right\rangle }
\newcommand{\bra}[1]{ \langle {#1} | }
\newcommand{\ket}[1]{ | {#1} \rangle }
\newcommand{\ev}{ {\rm eV} }
\newcommand{\kev}{{\rm keV}}
\newcommand{\mev}{{\rm MeV}}
\newcommand{\gev}{{\rm GeV}}
\newcommand{\tev}{{\rm TeV}} 
\newcommand{\mpl}{$M_{Pl}$}
\newcommand{\mw}{$M_{W}$}
\newcommand{\Ft}{F_{T}}
\newcommand{\Zparity}{\mathbb{Z}_2}
\newcommand{\BLambda}{\boldsymbol{\lambda}}
\newcommand{\be}{\begin{eqnarray}}
\newcommand{\ee}{\end{eqnarray}}
\newcommand{\met}{\;\not\!\!\!{E}_T}
\newcommand{\draftnote}[1]{{\bf\color{blue} #1}}

\title{{\bf Probing the Higgs Couplings to Photons in $h\rightarrow 4\ell$ at the LHC}}
\author{\bf{Yi Chen$\,^{a,\dagger}$,~Roni Harnik$\,^{b,\ddag}$,\,~Roberto Vega-Morales$\,^{b,c,d,\#}$}}

\affiliation{
$^a$Lauritsen Laboratory of Physics, California Institute of Technology, Pasadena, CA, 92115,\\
$^b$Theoretical Physics Department, Fermilab, P.O.~Box 500, Batavia, IL 60510, USA,\\
$^c$Laboratoire de Physique Th\'{e}orique d'Orsay, UMR8627-CNRS, Universit\'{e} Paris-Sud, Orsay, France,\\
$^d$Department of Physics and Astronomy, Northwestern University, Evanston, IL 60208, USA}

\email{$^\dagger$yichen@caltech.edu\\
$^\ddag$roni@fnal.gov\\
$^\#$roberto.vega@th.u-psud.fr}

\begin{abstract}

We explore the sensitivity of the Higgs decay to four leptons, the so-called golden channel, to higher dimensional loop-induced couplings of the Higgs boson to $ZZ$, $Z\gamma$, and $\gamma\gamma$, allowing for general CP mixtures.~The larger standard model tree level coupling $hZ^\mu Z_\mu$ is the dominant ``background'' for the loop induced couplings.~However this large background interferes with the smaller loop induced couplings, enhancing the sensitivity.~We perform a maximum likelihood analysis based on analytic expressions of the fully differential decay width for $h\to 4\ell$ ($4\ell \equiv 2e2\mu, 4e, 4\mu$) including all interference effects.~We find that the spectral shapes induced by Higgs couplings to photons are particularly different than the $hZ^\mu Z_\mu$ background leading to enhanced sensitivity to these couplings.~We show that even if the $h\to\gamma\gamma$ and $h\to 4\ell$ rates agree with that predicted by the Standard Model, the golden channel has the potential to probe both the CP nature as well as the overall sign of the Higgs coupling to photons well before the end of high-luminosity LHC running~($\sim$3 ab$^{-1}$).

\end{abstract}

\preprint{FERMILAB-PUB-14-048-PPD-T, LPT-Orsay-14-13, CALT 68-2883, nuhep-th/14-02}

\maketitle

\section{Introduction}
\label{sec:Intro}

With the recent discovery of the Higgs boson at the LHC~\cite{:2012gk,:2012gu} the focus now shifts to the determination of its detailed properties and in particular whether or not it possesses any anomalous couplings not predicted by the Standard Model (SM).~The Higgs decay to electrons and muons through electroweak gauge bosons, the so called golden channel, has been well established as a means towards accomplishing this goal as evidenced by the many studies of this channel~\cite{Nelson:1986ki,Soni:1993jc,Chang:1993jy,Barger:1993wt,Arens:1994wd,Choi:2002jk,Buszello:2002uu,Godbole:2007cn,Kovalchuk:2008zz,Cao:2009ah,Gao:2010qx,DeRujula:2010ys,Gainer:2011xz,Coleppa:2012eh,Bolognesi:2012mm,Boughezal:2012tz,Belyaev:2012qa,Avery:2012um,Campbell:2012ct,Campbell:2012cz,Grinstein:2013vsa,Modak:2013sb,Sun:2013yra,Gainer:2013rxa,Chen:2013waa,Anderson:2013fba,Buchalla:2013mpa,Gainer:2014hha}.~Various methods were established to probe the Higgs couplings to $ZZ$ pairs motivating experimental studies of their CP properties~\cite{Chatrchyan:2012jja,Chatrchyan:2013mxa,CMS-PAS-HIG-14-014, Khachatryan:2014kca} where CP odd/even mixtures as large as $\sim40\%$ are found to still be allowed.~However, apart from recent studies~\cite{Stolarski:2012ps,Chen:2012jy,Chen:2013ejz,Chen:2014pia}, the potential for the $h\to 4\ell$ ($4\ell \equiv 2e2\mu, 4e, 4\mu$) decay to probe the Higgs couplings to $Z\gamma$ and $\gamma\gamma$ pairs (we do not distinguish between on/off-shell vector bosons) has largely been neglected.

It is typically thought that these contributions are too small to be detected in the golden channel since they only first occur at loop level with the photon forced to be off-shell.~The search for these couplings is thus done solely using the rates of the decays $h\to Z\gamma$~and~$\gamma\gamma$ respectively.~In this note we show that large differences in shapes of the kinematic distributions allow for the possibility of measuring these couplings in the golden channel even if no significant deviations from the SM prediction are seen in the overall decay rates of $h\to\gamma\gamma$ or $h\rightarrow 4\ell$.~Furthermore, the various interference effects, in particular those with the tree level SM $hZ^\mu Z_\mu$ operator, also allow for the CP properties of these couplings to be extracted.~To demonstrate this we treat the SM $hZ^\mu Z_\mu$ operator as a `background' and perform a simple analysis to assess to what extent the golden channel has or will have any sensitivity to the various possible loop induced couplings of the Higgs boson to not only $ZZ$, but also $Z\gamma$ and $\gamma\gamma$ pairs for values of couplings that are of the same order as those predicted by the SM at one loop.

As we will see, the greatest sensitivity will turn out to be for the $\gamma\gamma$ couplings.~In fact we will show that the sensitivity is strong enough that there are excellent prospects for establishing the CP properties and overall sign of the Higgs couplings to photons during LHC running.~This is particularly exciting because in order to measure the CP properties of this coupling in the $h\rightarrow \gamma\gamma$ decay one would need to measure the polarization of the final state photons at such energies~\cite{Bishara:2013vya}.~Although there are recent proposals for making such a measurement~\cite{Voloshin:2012tv,Bishara:2013vya} using photon pair conversions, they are experimentally very challenging and it is not yet clear if they are viable options at the LHC.

There are also indirect approaches such as using measurements of the electric dipole moment~\cite{McKeen:2012av,Baron:2013eja} which place severe constraints on the amount of CP violation allowed in the Higgs to photon couplings.~Even these however rely on (perfectly reasonable) theory assumptions, such as what the scale of new physics is, and that the Higgs has direct couplings to light fermions.~Such couplings to light fermions have yet to be measured and perhaps will not be for some time.~In fact, it was shown that these constraints can be evaded completely in simple specific models~\cite{McKeen:2012av}.~It would thus be satisfying to have a direct probe of these couplings which does not rely on such assumptions.~Furthermore, these measurements are unable to determine the overall sign (or phase) of the effective Higgs $\gamma\gamma$ coupling.~This makes the golden channel perhaps the unique method of \emph{directly} probing the CP properties, including the overall sign, of the Higgs couplings to photons with current experimental technology and without theoretical assumptions.

Using a maximum likelihood analysis based on an analytic framework developed in~\cite{Chen:2013ejz}, we perform a simultaneous parameter extraction of the loop induced $ZZ$, $Z\gamma$, and $\gamma\gamma$ effective Higgs couplings allowing for general CP odd/even mixtures.~We perform these fits for a range of numbers of events assuming a pure SM data set.~We find that for values of couplings close to those predicted by the SM, the golden channel has excellent prospects to begin directly probing the Higgs couplings to photons during LHC running with $\sim 100-400 fb^{-1}$ of luminosity (depending on detector performance and production uncertainties) with less optimistic prospects for the $Z\gamma$ and even less so for the loop induced $ZZ$ couplings.~Our analysis is done at generator level neglecting any detector effects as well as any backgrounds but as we discuss further below, this is not expected to affect our results dramatically or change our conclusions qualitatively~\cite{Chen:2013ejz,Chen:2014pia}.

The results presented here motivate a detailed loop analysis in order to make more precise quantitative statements about the ability to extract these parameters.~They also suggest exciting potential for the golden channel to discover new physics which may enter in the loops that generate these effective couplings.~We leave a careful study of these issues to ongoing and future work~\cite{Chen:2014pia,followup2}.~For now we simply demonstrate qualitatively that the LHC has excellent prospects to establish the CP nature of the Higgs couplings to photons, including the overall sign, well before the end of high luminosity LHC running.
 ($\sim 3~ab^{-1}$).
 
This paper is organized as follows: In Secs.~\ref{sec:goldenchannel} we discuss the parameterization of the various tensor couplings which we will be fitting for as well as other aspects of searching for anomalous couplings with the golden channel.~In Sec.~\ref{sec:results} we present our results where we estimate the expected sensitivity of the golden channel to each of the loop induced effective Higgs couplings to $ZZ$, $Z\gamma$, and $\gamma\gamma$ pairs.~Finally in Sec.~\ref{sec:conclusion} we discuss briefly ongoing and future work before concluding.

\section{Examining the Golden Channel}
\label{sec:goldenchannel}
In this section we examine various aspects of the golden channel.~We begin by parametrizing the Higgs couplings to $ZZ$, $Z\gamma$, and $\gamma\gamma$ pairs.~We then discuss some of the observables which enable us to have sensitivity to these couplings and the different terms which contribute to the differential cross section.~We also examine the magnitude of the effects of loop induced couplings and discuss the interference effects.
\\
\subsection{Higgs Couplings to EW Bosons}
\label{subsec:couplings}
We consider the leading contributions to the Higgs couplings to neutral electroweak gauge bosons allowing for general CP odd/even mixtures as well as for $ZZ$, $Z\gamma$ and $\gamma\gamma$ to contribute simultaneously.~These couplings are parametrized by the following Lagrangian,
\begin{widetext}
\begin{eqnarray}
\label{eqn:siglag}
\mathcal{L} &\supset& \frac{h}{4v} 
\Big(2 A_{1}^{ZZ} m_Z^2Z^\mu Z_\mu + A_{2}^{ZZ} Z^{\mu\nu}Z_{\mu\nu} + A_{3}^{ZZ} Z^{\mu\nu} \widetilde{Z}_{\mu\nu} \nonumber\\
&+&~2A_{2}^{Z\gamma} F^{\mu\nu}Z_{\mu\nu} +2A_3^{Z\gamma} F^{\mu\nu} \widetilde{Z}_{\mu\nu}  
+~A_2^{\gamma\gamma} F^{\mu\nu}F_{\mu\nu} + A_3^{\gamma\gamma} F^{\mu\nu} \widetilde{F}_{\mu\nu} \Big) ,
\end{eqnarray}
\end{widetext}
where we have taken $h$ real.~We consider only up to dimension five operators and $Z_\mu$ is the $Z$ field while $V_{\mu\nu} = \partial_\mu V_\nu - \partial_\nu V_\mu$ is the usual bosonic field strengths.~The dual field strengths are defined as $\widetilde{V}_{\mu\nu} = \frac{1}{2} \epsilon_{\mu\nu\rho\sigma} V^{\rho \sigma}$.~All of the couplings are taken to be real\footnote{Our framework can easily accommodate complex couplings, but we expect any phases to be small~\cite{Bishara:2013vya} and their inclusion is not necessary in order to make our point.}, dimensionless, and constant.~In principal they are form factors whose loop functions have potentially strong momentum dependence due to the highly off-shell nature of the intermediate vector bosons.~This is true even in the SM where at tree level the only contribution is $A_1^{ZZ}$, but at one loop momentum dependent form factors of $\mathcal{O}(10^{-2} - 10^{-3})$ are generated for the $A_{2}^{ZZ, Z\gamma, \gamma\gamma}$ operators~\cite{Bredenstein:2006rh,Bredenstein:2006nk} by loops of SM particles ($A_{3}^{ZZ, Z\gamma, \gamma\gamma}$ are also generated at higher loop order, but these are totally negligible in comparison).

However, since we only aim to give a qualitative picture of the sensitivity and not a precise extraction of these parameters, for the purposes of this study we work within Higgs effective theory and approximate the couplings as constant, as is done in other similar analyses~\cite{Gao:2010qx,Bolognesi:2012mm,Anderson:2013fba,Avery:2012um,Artoisenet:2013puc,Chen:2012jy,Chen:2013ejz,Chen:2014pia}.~Once sensitivity of $\mathcal{O}(10^{-2} - 10^{-3})$ is achieved a more precise quantification will require accounting for the full momentum dependence, but we leave this to future work.~Thus for the remainder of this study we define as the SM point $A_1^{ZZ} = 2$ and take all other couplings $\sim 0$.~The purpose of this study is then to estimate at what point the golden channel will reach sensitivities of $\mathcal{O}(10^{-2} - 10^{-3})$ to the loop induced couplings assuming the `true' value of these couplings is that predicted by the SM (or close to it).~Achieving this level of sensitivity is exciting not only because one would begin probing SM loop effects, but also because of the potential for discovering new physics in deviations from the SM expectation including the possibility of observing CP violation.

\subsection{The Fully Differential Decay Rate}
\label{subsec:fullydiffw}

We have analytically computed and validated the fully differential decay width for $h\to 4\ell$ for the $2e2\mu$~\cite{Chen:2012jy}, $4e$, and $4\mu$~\cite{Chen:2013ejz} final states assuming on-shell decay of the Higgs.~All interference effects between the operators in Eq.(\ref{eqn:siglag}) as well as identical final state interference in the case of $4e$ and $4\mu$ have been included.

For the purpose of our analysis it is useful to note that the fully differential decay width for $h\rightarrow 4\ell$ is a sum over terms quadratic in the couplings which we can write schematically as,
\begin{eqnarray}
\label{eqn:diffwidth}
\frac{d\Gamma_{h\rightarrow 4\ell}}{d\mathcal{O}} 
\sim \sum A_n^{i} A_{m}^{j\ast} 
\times \frac{d\hat\Gamma_{nm}^{ij}}{d\mathcal{O}},
\end{eqnarray}
where the sum is over $n,m = 1,2,3$ and $i,j = ZZ, Z\gamma,\gamma\gamma$ (note $A_1^{Z\gamma} = A_1^{\gamma\gamma} = 0$).~We also define $d\mathcal{O} = dM_1^2dM_2^2d\vec{\Omega}$ which represents the differential volume element, or phase space, in terms of two invariant masses corresponding to the two lepton pairs ($M_1, M_2$) and five angles ($\vec{\Omega}$)~\cite{Chen:2013ejz,Chen:2014pia} (two of the angles correspond to an overall rotation in the Higgs frame and are not useful for resolving Higgs couplings).~It will also be useful to define,
\begin{equation}
\label{eqn:subrate}
\frac{d\Gamma_{nm}^{ij}}{d\mathcal{O}}\equiv A_n^{i} A_{m}^{j\ast} 
\times \frac{d\hat\Gamma_{nm}^{ij}}{d\mathcal{O}}.
\end{equation}
These differential distributions are going to be helpful in the following discussion and it will be convenient to give them names.~As they sum up to give the differential rate, we will call $d\Gamma^{ij}_{nm}/d\mathcal{O}$ a differential \emph{sub-rate}.

It is further useful to note that sub-rates fall into two categories - squared terms for which $n=m$ and $i=j$, and the interference terms for which this is not the case.~The squared terms are positive definite, while the interference terms may be both positive and negative.~In fact, the interference terms between CP odd and CP even operators must integrate to zero over all of phase space and thus must take on both signs.

\subsection{Sub-leading Couplings in $h\to4\ell$:\\ Yesterday's Signal = Today's Background} 
\label{subsec:bgandsignal}

In this work we focus on the question - to what degree is $h\to 4\ell$ sensitive to the small higher-dimensional couplings~$A_{2,3}^{ZZ}$, ~$A_{2,3}^{Z\gamma}$, and~$A_{2,3}^{\gamma\gamma}$ in Eq.(\ref{eqn:siglag})?~For the purpose of this question we can think of the dominant decay $h\to ZZ \to 4\ell$ via $A_1^{ZZ}$ as the \emph{background}.\footnote{Of course, there are also true SM backgrounds contributing to $pp\to4l$, but these are relatively small and will not change our results qualitatively~\cite{Chen:2012jy,Chen:2013ejz,Chen:2014pia}.~They are henceforth ignored in this work but should be part of a more complete analysis.}~The small deviations in the fully differential cross-section caused by the presence of higher dimensional operators $A_{2,3}^{i}$ are our \emph{signal}.~The various signal sub-rates are correlated in the sense that turning on a squared sub-rate inevitably leads to the presence of an interference term, and vice versa.

There are a few qualitative factors that are important in determining whether the differential $h\to 4\ell$ rate is sensitive to the loop induced couplings, i.e.~our signal:
\begin{itemize}
\item \emph{Shapes:}~To what degree do the multi-dimensional distributions $d\Gamma_{nm}^{ij}/d\mathcal{O}$ in equation~(\ref{eqn:subrate}) for our signal differ from the distribution predicted by the $A_1^{ZZ}$ background?~A bigger difference in shape will lead to better sensitivity.
\item\emph{The total sub-rates:}~Irrespective of the size of the couplings $A_{n}^{i}$, what are the sizes of the differential sub-rates $d\hat\Gamma_{nm}^{ij}/d\mathcal{O}$ which contribute to the total $h\to 4\ell$ differential decay rate?~Sensitivity will be enhanced to couplings with larger sub-rates.
\item\emph{Interference:}~Given particular values for the loop induced $A_{n}^{i}$ couplings, are the dominant signal sub-rates coming from one of the squared terms or from interference?~This will determine the progression of the sensitivity with the growing luminosity since the former is quadratic in the coupling while the latter is linear.
\end{itemize} 
In the next subsections we will consider these factors in more detail.~First we examine the shapes of the differential sub-rate with respect to the invariant masses.~Then we will examine the size of the various sub-rates before discussing the interference effects.~These studies will help us understand the results of our full likelihood analysis and the progression of sensitivity with luminosity which will be shown in Sec~\ref{sec:results}.

\subsection{The Differential Mass Spectra} 
\label{subsec:distributions}
The power of the golden channel comes from the large number of observables available in the $4\ell$ final state and their correlations which provide a vast amount of information.~Focusing on only decay observables and taking the Higgs mass as input, we have the two invariant masses, corresponding to the two lepton pairs, and three angles of relevance as discussed above (see~\cite{Chen:2013ejz,Chen:2014pia} for more details).~The shapes of the distributions, $d\Gamma_{nm}^{ij}/d\mathcal{O}$ in Eq.(\ref{eqn:subrate}), are in general quite different for the various $ZZ$, $Z\gamma$ and $\gamma\gamma$ contributions allowing for strong discriminating power between the different possible operators.~This means that even if the overall $h\rightarrow 4\ell$ rate is consistent with the SM prediction one can still have contributions from new physics which can be uncovered in the differential distributions.~In terms of the SM this translates to saying that even though $A_{2,3}^{i}$ contribute negligibly to the overall rate in principal the golden channel still can have sensitivity to these loop induced couplings.

\begin{figure}[t]
\includegraphics[width=.47\textwidth]{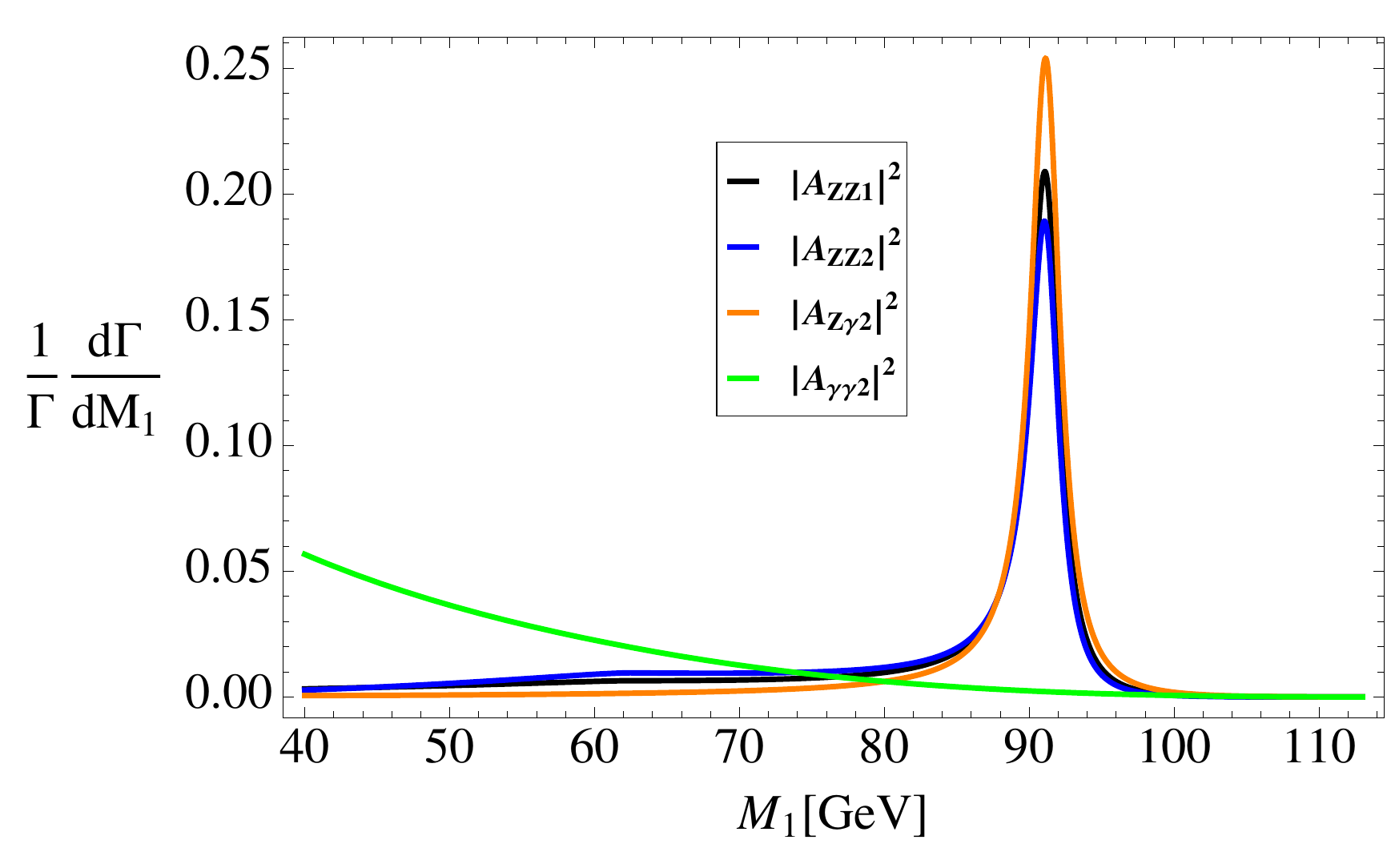}
\includegraphics[width=.47\textwidth]{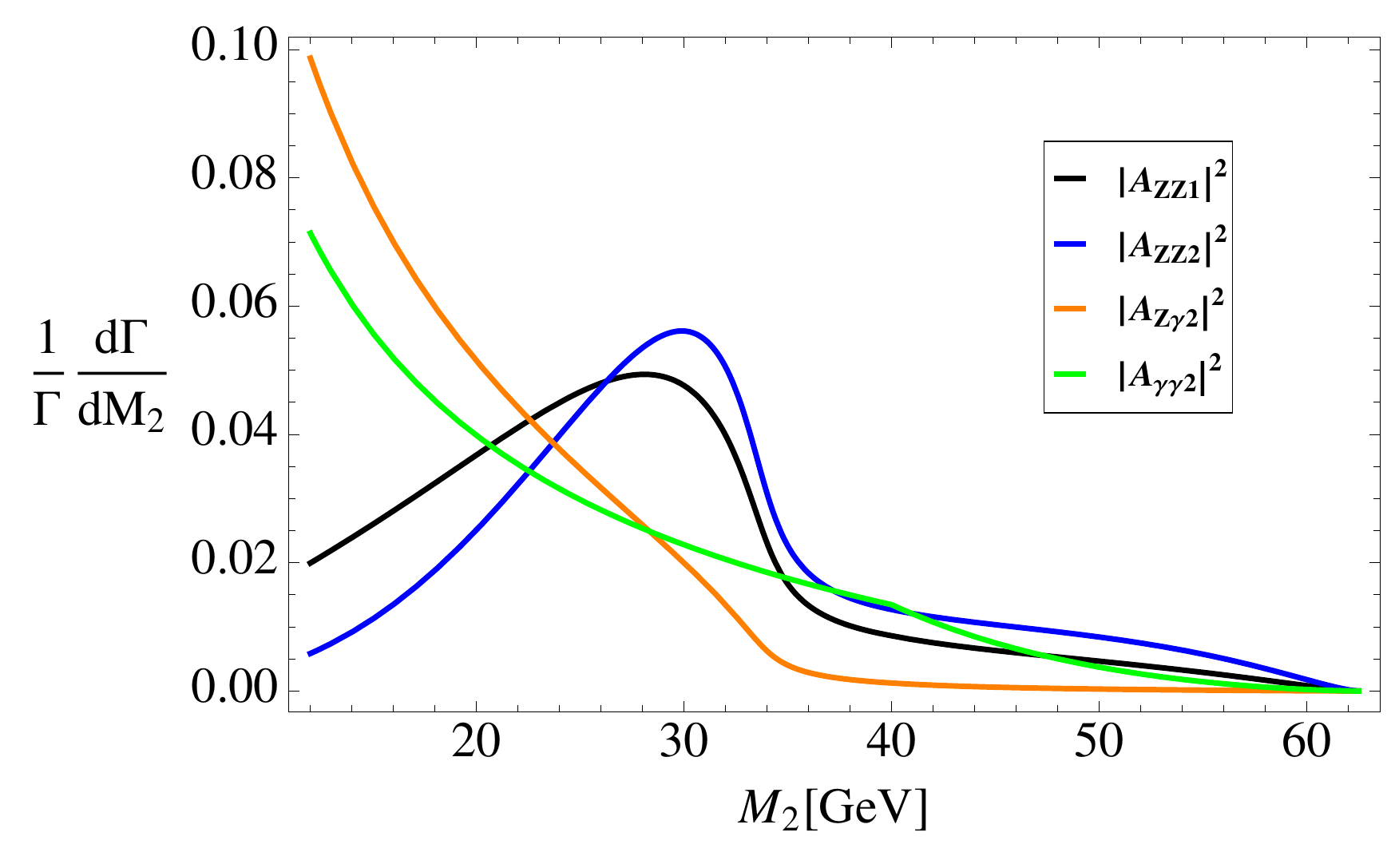}
\caption{{\bf Top:}~The differential mass spectrum for $M_1$ in the $2e2\mu$ final state for the CP even terms squared plotted on top of the SM `background' shown in black.~{\bf Bottom:}~The differential mass spectrum for $M_2$ in the 2$e2\mu$ final state for the same combination of operators.~To obtain the spectra we have performed the integration over the full angular phase space analytically and restricted to the range $50 \leq M_1 \leq 120~$GeV and $12 \leq M_2 \leq 60~$GeV with no other cuts and normalized them to one.}
\label{fig:M1M2squared}
\end{figure}

We can get a sense for the shape differences in these distributions by examining the differential spectra for the two invariant masses obtained via integration of Eq.(\ref{eqn:subrate}) over all angles and one invariant mass.~The invariant masses serve as the most strongly discriminating variables between the different operators.~By examining their distributions we can thus get a good qualitative picture of the relative sensitivity to the various operators.~These are presented in Fig.~\ref{fig:M1M2squared} for the $2e2\mu$ final state where we show the distributions for the invariant mass which reconstructs closest to the $Z$ mass which we call $M_1$ and the `off-shell' invariant mass which we call $M_2$.~We show the distributions (all normalized to one) for the four CP even operators squared corresponding to $|A_1^{ZZ}|^2$, $|A_2^{ZZ}|^2$, $|A_2^{Z\gamma}|^2$, and $|A_2^{\gamma\gamma}|^2$.~One can see that for both the $M_1$ and $M_2$ distributions, the shape of $|A_2^{\gamma\gamma}|^2$ (green) is the operator most easily distinguished from the $|A_1^{ZZ}|^2$ `background' (black).~The next most distinguishable operator, mostly in $M_2$, is $|A_2^{Z\gamma}|^2$ (orange) followed by $|A_2^{ZZ}|^2$ (blue) which as expected most closely resembles the $|A_1^{ZZ}|^2$ background.~The shapes for the CP odd squared terms follow a similar pattern and are thus not shown.~Though we do no show it here, we note that the relative azimuthal angle between the lepton decay planes is useful for resolving CP even and CP odd operators~\cite{Cao:2009ah,Chen:2012jy} (as is $M_2$~\cite{Boughezal:2012tz,Chen:2012jy}).

\subsection{Interference in the Golden Channel} 
\label{subsec:interference}

\begin{figure}[t]
\includegraphics[width=.47\textwidth]{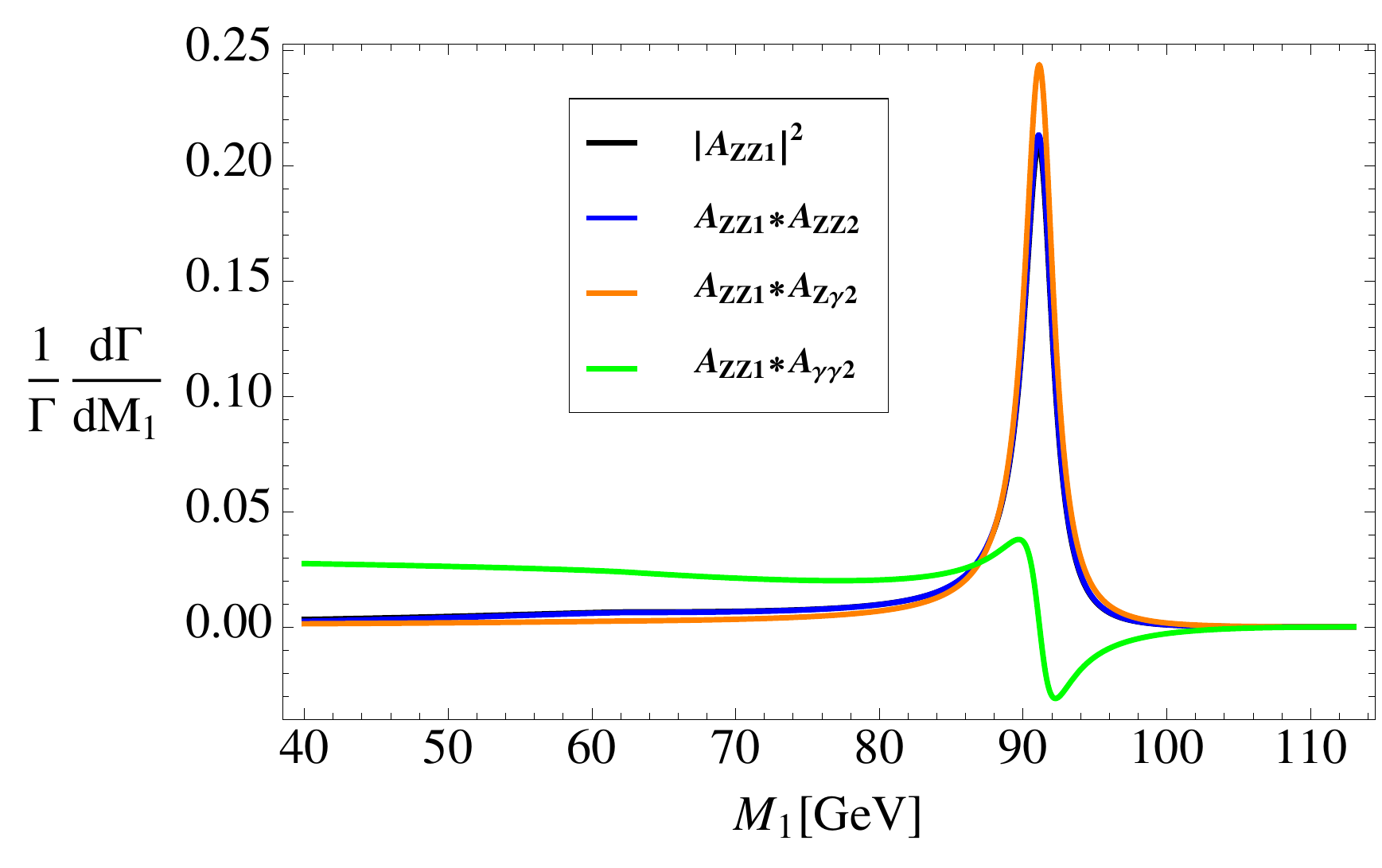}
\includegraphics[width=.47\textwidth]{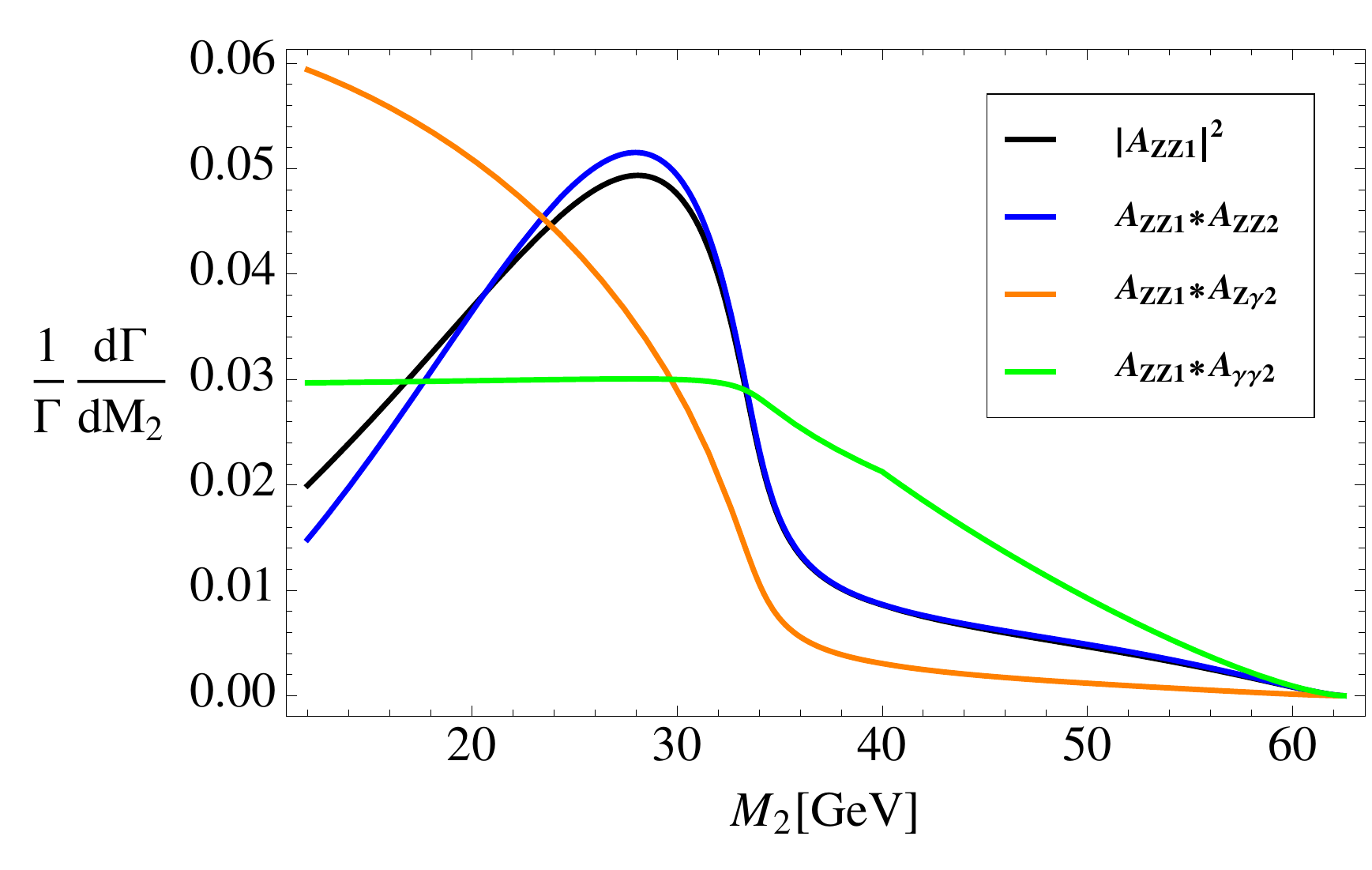}
\caption{{\bf Top:}~The differential mass spectrum for $M_1$ in the $2e2\mu$ final state for the CP even interference with the SM operator $A_1^{ZZ}$.~The SM `background' is again shown in black.~{\bf Bottom:}~The differential mass spectrum for $M_2$ in the 2$e2\mu$ final state for the same combination of operators.~The cuts are identical to those in Fig.~\ref{fig:M1M2squared} and we have normalized the distributions to positive one.}
\label{fig:M1M2interference}
\end{figure}
A second advantage of the golden channel is its sensitivity to interference which means that we are probing effects which are linear in $A_{2,3}^{i}$ in addition to the squared terms $|A_{2,3}^{i}|^2$.~This makes the golden channel sensitive to the CP properties of the couplings as well as their overall sign.~This is a feature not present, for example, in  $h\rightarrow \gamma\gamma$ rate measurements which are sensitive only to the combination $|A_{2}^{\gamma\gamma}|^2 + |A_{3}^{\gamma\gamma}|^2$~\cite{Bishara:2013vya}.~We can get a feel for the effects of this interference by again examining the $M_1$ and $M_2$ distributions, but this time for the interference between the higher dimensional operators $A^i_{2,3}$ and the tree level operator $A_1^{ZZ}$.~These are shown in Fig.~\ref{fig:M1M2interference}.~As mentioned previously these distributions can take on both positive and negative values.~For those shown in Fig.~\ref{fig:M1M2interference} we have normalized them such that when integrated over the invariant mass we obtain positive one.~Again we have plotted on top of the $|A_1^{ZZ}|^2$ background (black).~We see a similar pattern of discrimination as that found for the squared terms in the sense that $\gamma\gamma$ is most easily distinguished from the background followed by $Z\gamma$ and then $ZZ$.

We note that it is not just interference with the SM that is important to consider if one is to avoid a bias during the parameter extraction procedure.~The different squared terms as well as all other possible interference terms among the loop induced operators are also important to include.~This is especially true with small data sets where fluctuations in the data can be mistaken for large anomalous Higgs interactions including subtle interference effects between the various operators.~We can gain further insight of this behavior and the possible interference effects by examining the total size of each possible combination of operators which we now discuss.

\subsection{The Integrated Magnitudes}
\label{subsec:phasespace}

Upon integration of Eq.(\ref{eqn:subrate}) over the full phase space we can obtain the total sub rates for each combination of $A^i_n A^{j\ast}_m$ couplings as follows,
\begin{eqnarray}
\label{eqn:totw}
\Gamma_{nm}^{ij} = A_n^{i} A_{m}^{j\ast} \times \int \frac{d\hat\Gamma_{nm}^{ij}}{d\mathcal{O}} d\mathcal{O}
\end{eqnarray}
Thus we can think of $\Gamma_{nm}^{ij}$ as the total `decay width' for the corresponding pair of operators, though again these can be negative for certain combinations of operators and so are not strictly speaking total decay widths.~Some of these interference terms are exactly zero in the case where one has a CP even operator mixing with a CP odd operator, i.e.~CP violation.~This is just representative of the fact that the overall $h\rightarrow 4\ell$ rate is not sensitive to CP violation though of course this does not mean that the golden channel is not sensitive to this effect.

\begin{figure}[t]
\includegraphics[width=.49\textwidth]{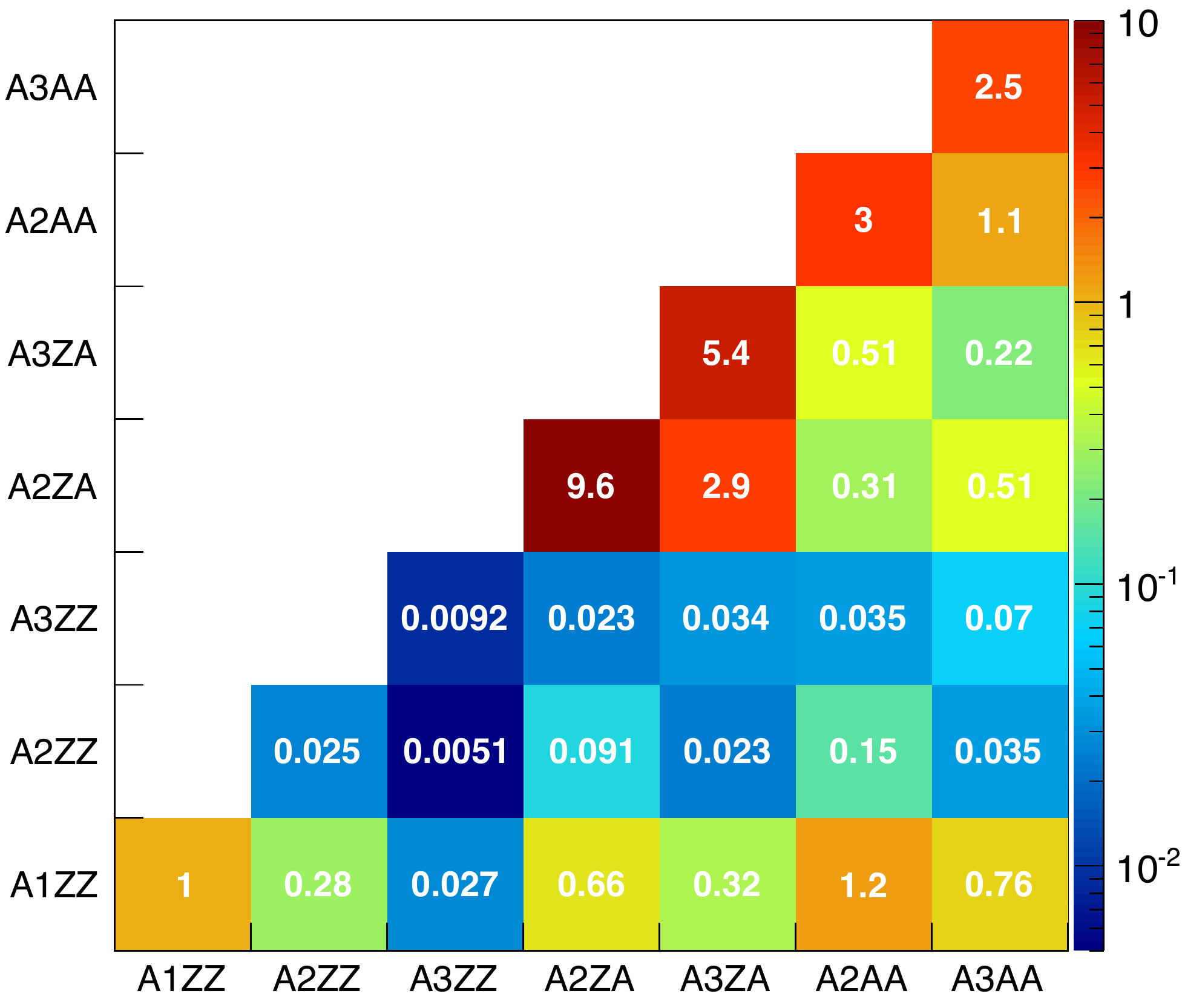}
\caption{The total integrated magnitudes, $\Pi^{ij}_{nm}$, defined in Eq.(\ref{eqn:absdiffw}), which correspond to the pairs of couplings $A^i_n A^{j\ast}_m$.~To obtain the values here we have set $A_1^{ZZ} = 2$ and all other couplings to one.~We have normalized to the (tree level) SM value for the $h\rightarrow 4\ell$ decay width.~The values shown are for the $2e2\mu$ final state~\cite{Chen:2013ejz} for a `CMS-like' phase space which is defined in Sec.~\ref{subsec:fitandPS}.~These magnitudes are useful for estimating the sensitivity in early stages of the analysis.}
\label{fig:CMS2e2muAbsMatrix}
\end{figure}

It is therefore more illuminating to show what we call the \emph{integrated magnitude} of the various combination of operators defined for each pair of couplings as,
\begin{eqnarray}
\label{eqn:absdiffw}
\Pi_{nm}^{ij} &=& A_n^{i} A_{m}^{j\ast} \times
\int \left| \frac{d\hat\Gamma_{nm}^{ij}}{d\mathcal{O}} \right| d\mathcal{O} ,
\end{eqnarray}
where the $\Pi_{nm}^{ij}$ are strictly non-zero even in the case of CP violation.~We show in Fig.~\ref{fig:CMS2e2muAbsMatrix} all possible combinations of $\Pi_{nm}^{ij}$ for $A_1^{ZZ} = 2$ and all loop induced couplings set to one.~We have normalized to the (tree level) SM value for the $h\rightarrow 4\ell$ decay width ($\Gamma_{4\ell}^{SM}$) which corresponds to $A_1^{ZZ} = 2$ and all other couplings set to zero.~The values shown are for $\Pi_{nm}^{ij}/\Gamma_{4\ell}^{SM}$ in the $2e2\mu$ final state~\cite{Chen:2013ejz} with cuts and reconstruction corresponding to a `CMS-like' phase space~\cite{:2012gu} which we have defined in Sec.~\ref{subsec:fitandPS}.~These integrated magnitudes contain information not only about the total phase space contribution of each combination of operators, but also about the differences in shape.~It is for this reason that one can have non-zero values even for combinations of operators which lead to CP violation.
\begin{figure}
\includegraphics[width=.49\textwidth]{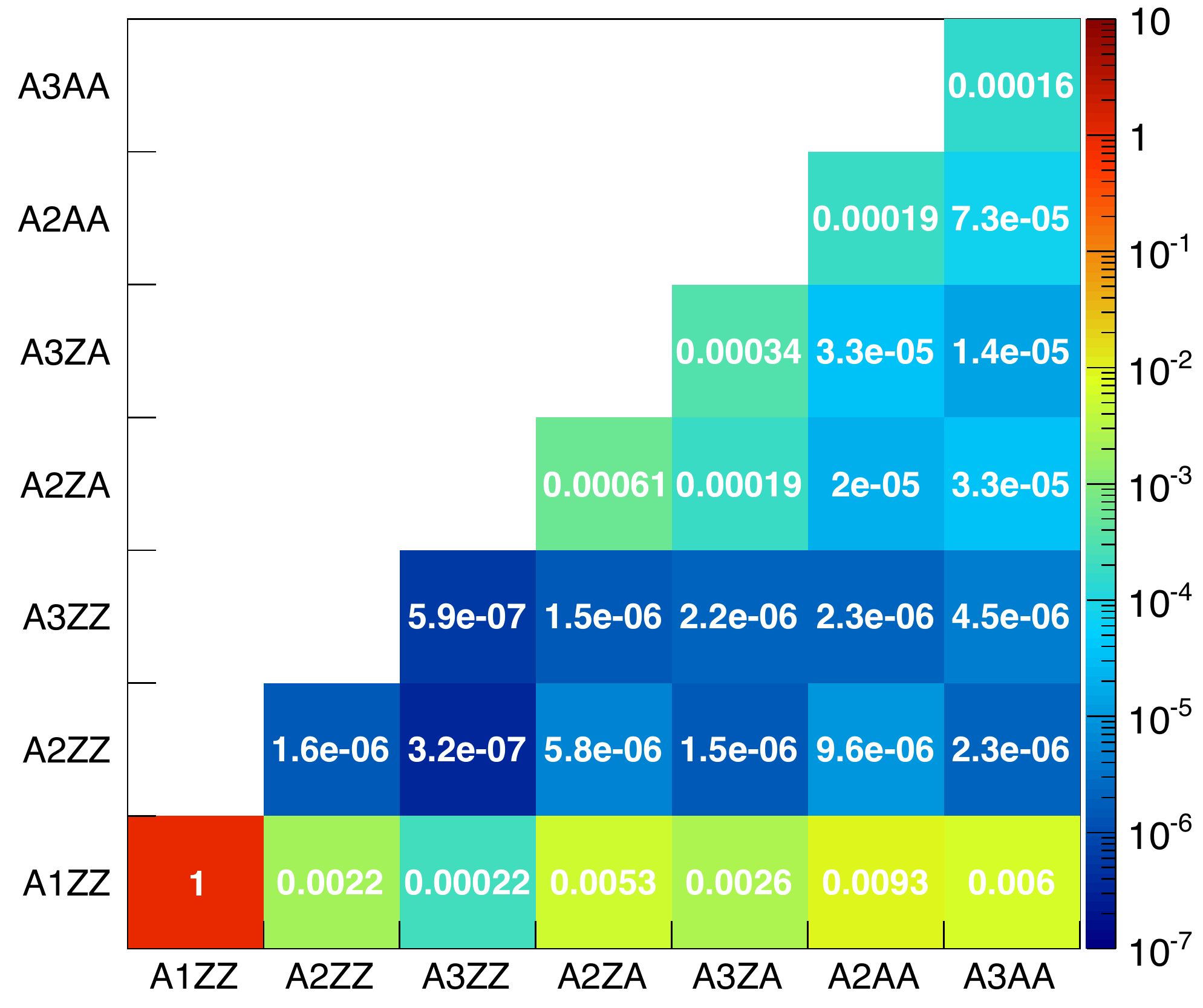}
\caption{The same as Fig.~\ref{fig:CMS2e2muAbsMatrix}, but with $A_1^{ZZ} = 2$ and all other couplings to $\sim 0.008$.~These values are useful to estimate the sensitivities of the various terms at late stages of LHC running.~We see that interference terms with the SM (first row) dominate over squared terms for all $A^i_{2,3}$.}
\label{fig:CMS2e2muAbsMatrix2}
\end{figure}

We can see by examining the diagonal terms that the largest integrated magnitudes are for the $Z\gamma$ and $\gamma\gamma$ contributions while the SM combination $|A_1^{ZZ}|^2$ is equal to one.~This is due to a combination of the fact that in these cases both gauge bosons can be more closely on-shell, as well as the larger coupling of photons to leptons relative to the $Z$ couplings.~These features contribute to the enhanced sensitivity to the $Z\gamma$ and $\gamma\gamma$ couplings as we will see in our results in Sec.~\ref{sec:results}.~In particular, this implies that generically we expect a greater sensitivity to the $Z\gamma$ and $\gamma\gamma$ couplings than for the $A_{2,3}^{ZZ}$ couplings, unless the $ZZ$ effective Higgs couplings, for some reason, are substantially larger than the $Z\gamma$ and $\gamma\gamma$ couplings.

The values in Fig.~\ref{fig:CMS2e2muAbsMatrix} were obtained for all loop induced couplings set equal to one.~Of course in the SM and in most new physics models we expect these couplings to be $\lesssim \mathcal{O}(10^{-2}-10^{-3})$ or much smaller.~We therefore again show $\Pi_{nm}^{ij}/\Gamma_{4\ell}^{SM}$ for the $2e2\mu$ final state in Fig.~\ref{fig:CMS2e2muAbsMatrix2}, but now with $A_1^{ZZ} = 2$ and all loop induced couplings set to $\sim 0.008$.~We see again that the SM combination $|A_1^{ZZ}|^2$ is equal to one (by definition).~Of the others, the interference terms between the signal operators and $A_1^{ZZ}$ dominate, with integrated magnitudes of $\sim 10^{-2}-10^{-3}$, and much smaller magnitudes for terms that involve two loop operators.~These small magnitudes may give the impression that there is no sensitivity in the golden channel to couplings other than $A_1^{ZZ}$ for parameter points `close to' the SM.~However as the discussion in previous sections indicates, one has much more information in the $h\to 4\ell$ fully differential decay width than just the integrated magnitudes.

From our discussions of the integrated magnitudes and differential spectra we naively expect that we should have the strongest sensitivity to the $\gamma\gamma$ couplings followed by the $Z\gamma$ couplings and the weakest sensitivity to the loop induced $ZZ$ couplings.~As we will show below, this indeed turns out to be the case.

\section{Results}
\label{sec:results}

To obtain our results we use the framework developed and described in detail in~\cite{Chen:2013ejz}.~We will take the SM tree level prediction of $A^{ZZ}_{1} = 2$ as input and fit to the remaining six couplings simultaneously.~Floating all parameters simultaneously ensures that we account for potentially important correlations between the various couplings~\cite{Chen:2013ejz}.~Note also that by fixing $A^{ZZ}_{1} = 2$ we are implicitly fitting to ratios of couplings and taking the overall normalization as input since it can be obtained from measurements of the total rate.~This also serves to minimize the dependence of our results on any production effects we have neglected.

For all of our results we combine the $2e2\mu$, $4e$, and $4\mu$ channels by computing the fully differential decay width for each final state~\cite{Chen:2012jy,Chen:2013ejz} (including identical final state interference for $4e$ and $4\mu$) and combining them into one likelihood.~The data sets which we fit to are generated from these expressions and contain a mixture of all three final states whose proportions are determined by the overall normalization of the differential widths for each channel.~Though we do not examine this issue here, we note that the three channels do not possess the same sensitivity.~We leave a detailed examination of this interesting point to an ongoing followup study~\cite{followup}.
\subsection{Fit and Phase Space Definition} 
\label{subsec:fitandPS}

We define our six dimensional parameter space as,
\begin{eqnarray}
\label{eqn:Aall}
\vec{A} = (A^{ZZ}_2,A^{ZZ}_3,A^{Z\gamma}_2,A^{Z\gamma}_3,A^{\gamma\gamma}_2,A^{\gamma\gamma}_3) .
\end{eqnarray}

To estimate the sensitivity we obtain what we call an `effective' $\sigma$ or \emph{average error} defined as~\cite{sigmadef},
\begin{eqnarray}
\label{eqn:sigma}
\sigma = \sqrt{\frac{\pi}{2}} \langle |\hat{A} - \vec{A}_o| \rangle ,
\end{eqnarray}
where $\hat{A}$ is the value of the best fit parameter point obtained by maximization of the likelihood with respect to~$\vec{A}$.~Here $\vec{A}_o$ represents the `true' value with which our data sets are generated.~The average error is then found by conducting a large number of pseudoexperiments with a fixed number of events and obtaining a distribution for $\hat{A}$ which will have some spread centered around the average value.~We then translate the width of this distribution into our effective $\sigma$ which converges to the usual interpretation of $\sigma$ when the distribution for $\hat{A}$ is perfectly gaussian.~We repeat this procedure for a range of fixed number of signal events to obtain $\sigma$ as a function of number of signal events $N_S$.

We take the Higgs mass to be $m_h = 125~$GeV and limit our phase space to approximate the cuts used by CMS as indicated by following cuts and reconstruction:
\begin{itemize}
	\item $p_{T\ell} > 20, 10, 7, 7~$GeV for lepton $p_T$ ordering,
	\item $|\eta_\ell| < 2.4$ for the lepton rapidity,
	\item $40$~GeV $\leq M_1$ and $12$~GeV $\leq M_2$. 
\end{itemize}
Here $M_1$ and $M_2$ are the reconstructed masses of the two lepton pairs.~In reconstructing $M_1$ and $M_2$ we always impose $M_1 > M_2$ and take $M_1$ to be the reconstructed invariant mass for a particle and anti-particle pair which is closer to the $Z$ mass.~Note however that two other lepton pairings are possible and equally valid, but we leave an exploration of these alternate reconstructions to ongoing work~\cite{followup}.~For further details on the fitting (maximization) procedure and on the statistical analysis see~\cite{Chen:2013ejz,Chen:2014pia}.


\subsection{Sensitivity as Function of Number of Events}
\label{subsec:fullfloatRMS}
Using the definition in Eq.(\ref{eqn:Aall}) we fit to a `true' parameter point,
\begin{eqnarray}
\label{eqn:Aall0}
\vec{A}_o = (0,0,0,0,0,0) ,
\end{eqnarray}
where we allow all six parameters in $\vec{A}$ to float simultaneously in the fit.~The `true' point $\vec{A}_o$ in Eq.(\ref{eqn:Aall0}) is roughly the prediction of the SM until getting to a precision of $\mathcal{O}(10^{-2} - 10^{-3})$ so it serves as a good `bench mark' point for us to estimate the sensitivity to the various couplings.
\begin{figure}[h]
\includegraphics[width=.52\textwidth]{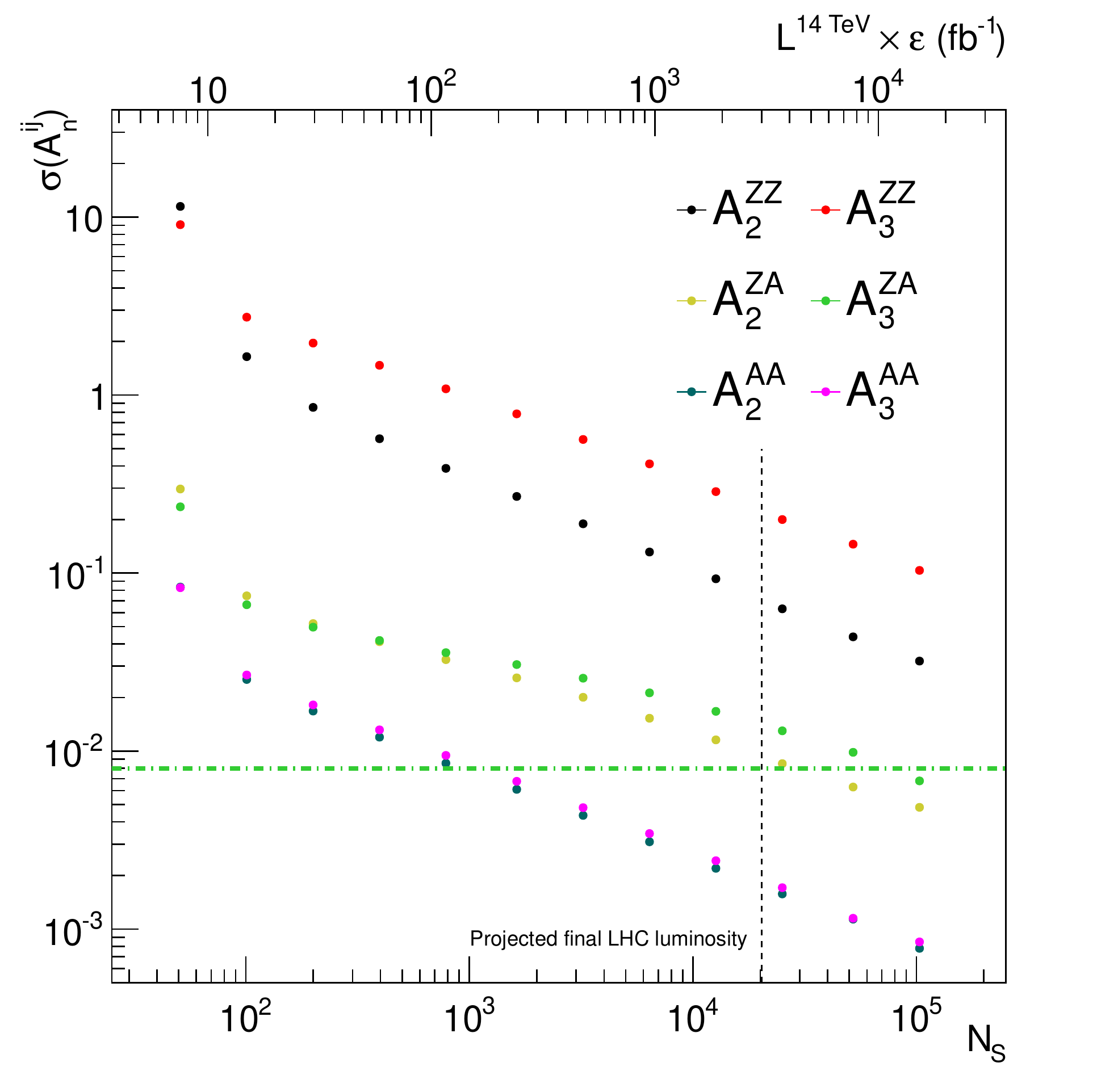}
\caption{The results for the effective $\sigma$, or average error defined in Eq.(\ref{eqn:sigma}), of each coupling as a function of the number of signal events $N_S$.~Error bars are shown, but they are smaller than the dot sizes.~We combine the $2e2\mu$, $4e$, and $4\mu$ channels in the data set and our likelihood.~On the top axis we also show an approximate projection for the luminosity $\times$ efficiency needed at the LHC to obtain a given number of signal events.~The vertical gray dashed line indicates a rough estimate for the final LHC luminosity which will be achieved ($\sim 3000 fb^{-1}$) using production cross section and branching fraction values obtained from the LHC Higgs Cross Section Working
Group~\cite{Dittmaier:2011ti,Heinemeyer:2013tqa}.~We indicate by the green dashed line the value $\sim 0.008$ corresponding roughly to the magnitude of $A_2^{\gamma\gamma}$ predicted by the SM at $125~$GeV.~All couplings are floated simultaneously and defined in Eq.(\ref{eqn:siglag}).}
\label{fig:LuminosityPlots}
\end{figure}
In Fig.~\ref{fig:LuminosityPlots} we show the result for $\sigma$~vs~$N_S$ for the six parameters defined in Eq.(\ref{eqn:siglag}).~Since all parameters are floated simultaneously these sensitivity projections include all correlations between the various couplings.~We indicate by the green dashed line the value $\sim 0.008$\footnote{This corresponds to the magnitude of $A_2^{\gamma\gamma}$ predicted by the SM at $125~$GeV for on-shell external photons~\cite{Low:2012rj}.} which we use as an approximate threshold for the necessary sensitivity to begin to probe these couplings in the SM.~On the top axis we also show an estimate for the expected LHC luminosity multiplied by efficiency while the vertical gray dashed line indicates a rough estimate for the final LHC luminosity which will be achieved ($\sim 3000 fb^{-1}$).~We have used production cross sections for both gluon fusion and vector boson fusion as well as the $h\rightarrow 4\ell$ branching fraction values provided by the LHC Higgs Cross Section Working Group~\cite{Dittmaier:2011ti,Heinemeyer:2013tqa}.

There are a number of interesting features to note in these results.~The first is the different slopes of the various sensitivity curves for each coupling.~These slopes can be understood by recalling the tables of the integrated magnitudes in Figs.~\ref{fig:CMS2e2muAbsMatrix} and~\ref{fig:CMS2e2muAbsMatrix2}.~Looking at the $Z\gamma$ curves we see a bending shape not seen for the other couplings.~This bending comes from the interplay between effects which dominate in two regimes.~One is when the squared terms drive the sensitivity (see Fig.~\ref{fig:CMS2e2muAbsMatrix}).~This occurs in the regime of smaller data sets when fluctuations lead to larger values of the couplings to be extracted in the maximization procedure, i.e.~larger errors.~Since the fluctuations of the `true' model (the SM with $A_1^{ZZ} = 2$ in this case) go like $N^{-1/2}$ this implies that when the squared terms dominate we expect $|A_{2,3}^{Z\gamma}|^2 \sim N^{-1/2}$ which means that the average error for the $Z\gamma$ couplings then scales like $\sigma(A_{2,3}^{Z\gamma}) \sim N^{-1/4}$.

The second regime occurs for larger data sets where smaller fluctuations allow for smaller values (closer to zero for the true SM point) of the loop induced couplings to be extracted.~Here we expect the interference terms with the SM to dominate (see Fig.~\ref{fig:CMS2e2muAbsMatrix2}).~Thus, now we have $A_1^{ZZ} \times A_2^{Z\gamma} = 2 \times A_2^{Z\gamma} \sim N^{-1/2} \Rightarrow \sigma(A_{2,3}^{Z\gamma}) \sim N^{-1/2}$.~The detailed shape of the curves will depend on where the transition from one regime to the other occurs.~For the $Z\gamma$ couplings this transition occurs later at larger event counts because of the large size of the squared terms to begin with (see Fig.~\ref{fig:CMS2e2muAbsMatrix}).~For the $ZZ$ and $\gamma\gamma$ couplings this transition occurs much earlier at smaller data sets and is therefore `hidden' in the highly non-gaussian region of low event count.~Thus for the $ZZ$ and $\gamma\gamma$ couplings the regime of $\sigma(A_{2,3}^{ZZ,\gamma\gamma}) \sim N^{-1/2}$ begins much sooner and we have the scaling observed in Fig.~\ref{fig:LuminosityPlots}.~Note however, that these considerations only describe the dominant behavior and the precise shape in the end is determined by the net effect of all possible contributions.

The next feature to note in Fig.~\ref{fig:LuminosityPlots} is that the sensitivity to the $\gamma\gamma$ couplings is significantly greater than for $Z\gamma$ and even more so than for $ZZ$.~This was to be expected from our considerations of the differential spectra in Sec.~\ref{subsec:distributions} as well as integrated magnitudes defined in Eq.(\ref{eqn:absdiffw}).~In fact we see that for the $\gamma\gamma$ couplings, $\sigma(A_{2,3}^{\gamma\gamma})$ reaches values $\lesssim \mathcal{O}(10^{-2})$ at around $\gtrsim 800$ events which corresponds to roughly $100 fb^{-1}$ of luminosity assuming $100\%$ efficiency.~Of course in reality the efficiency is much lower than this and there are large uncertainties on the production cross section, but conservatively we estimate this number of events can be reached with $\sim 300-400 fb^{-1}$ after accounting for detector and production effects.~The level of precision reached in the $\gamma\gamma$ couplings with $\sim800$ events starts to become of the same order or smaller than the $\mathcal{O}(10^{-2}-10^{-3})$ loop effects which generate $A_2^{\gamma\gamma}$ in the SM.~It is quite remarkable that the LHC will likely reach this level of sensitivity in the Higgs couplings to photons, perhaps even before a high luminosity upgrade.

Of course it is around this level of precision that loop momentum effects start to become important.~The off-shell nature of the intermediate vector bosons means these loop momentum effects could be sizable when approaching this level of precision.~In addition there are detector and production effects which must be accounted for before a more precise quantification can be given.~The framework for achieving this task has been developed in~\cite{Chen:2014pia,Chen:2014hqs}.

\subsection{Establishing the $h\gamma\gamma$ CP Properties}
\label{subsec:A2vA3}
The results shown in Fig.~\ref{fig:LuminosityPlots} indicate that the golden channel should be able to establish the CP nature and overall sign of the Higgs couplings to photons for couplings roughly of the same size as those predicted by the SM.~To explore this further we perform a second parameter extraction~This time to the `true' point given by,
\begin{eqnarray}
\label{eqn:A2AA}
\vec{A}_o = (0,0,0,0,-0.008,0) ,
\end{eqnarray}
again allowing all couplings to float.~We have chosen $A_2^{\gamma\gamma} = -0.008$ which is the value of predicted by the SM at $125~$GeV for on-shell external photons~\cite{Low:2012rj}.~Though in the golden channel the photons are off-shell we consider this a sufficient approximation for present purposes.

\begin{figure}[t]
\includegraphics[width=.48\textwidth]{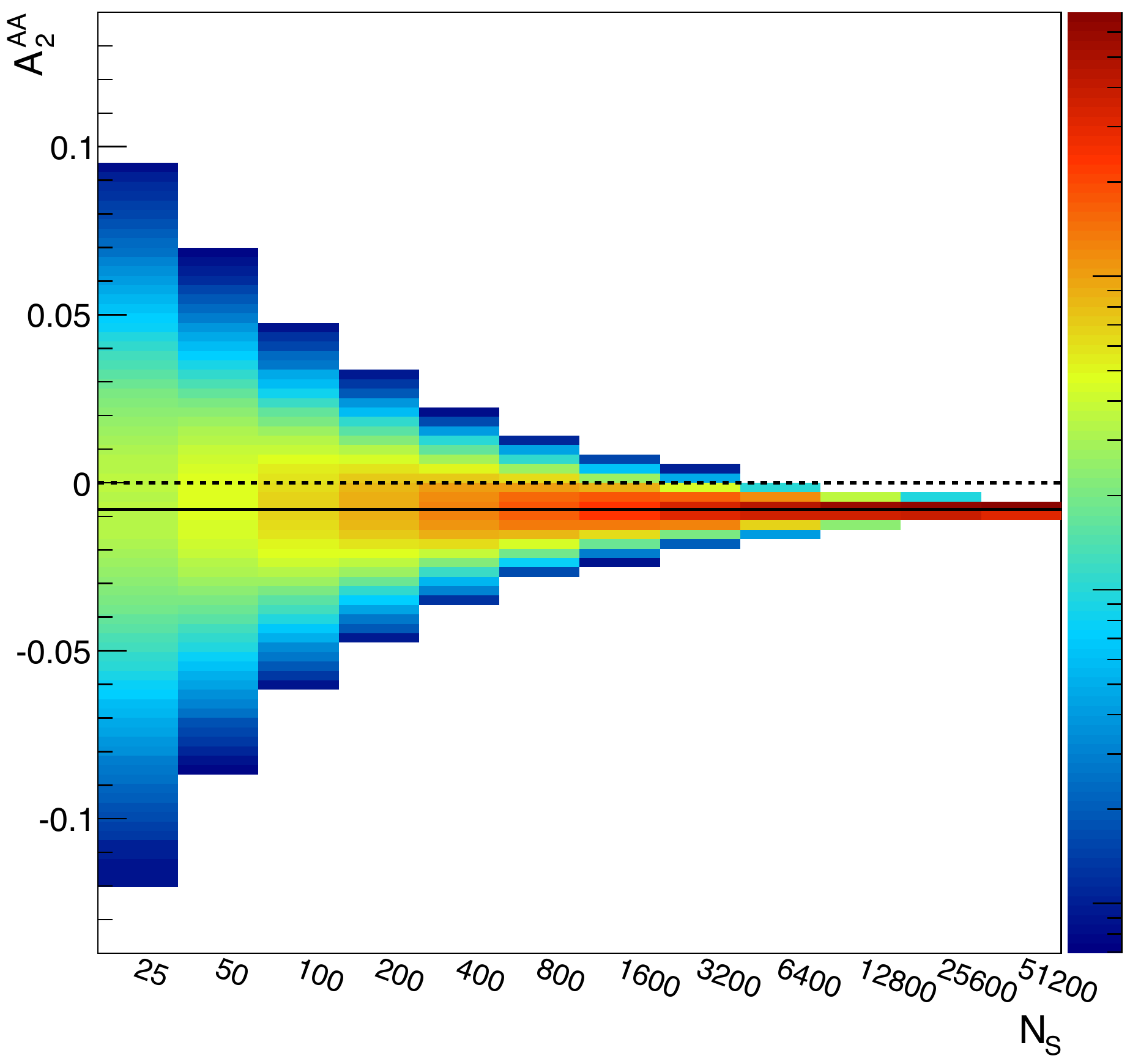}
\caption{The distribution of extracted $A_2^{\gamma\gamma}$ as a function of the number of events for 25 to 51200 signal events per pseudoexperiment for the `true' point defined in Eq.(\ref{eqn:A2AA}).~The true value $A_2^{\gamma\gamma} = -0.008$ is shown by the black solid line while zero is shown by the dashed black line.~The colors indicate the density of pseudoexperiments which return a particular value of $A_2^{\gamma\gamma}$ with red being high density and blue being low.~Note that although we only show $A_2^{\gamma\gamma}$, all parameters in Eq.(\ref{eqn:Aall}) are floated in the fit.~Again we combine the $2e2\mu$, $4e$, and $4\mu$ channels in the data set and our likelihood.}
\label{fig:coolplot}
\end{figure}

We first show in Fig.~\ref{fig:coolplot} results for the distribution of extracted $A_2^{\gamma\gamma}$ as a function of the numbers of events for 25 to 51200 signal events per pseudoexperiment.~Although we only show the distribution for $A_2^{\gamma\gamma}$, all parameters in Eq.(\ref{eqn:Aall}) are floated in the fit and thus Fig.~\ref{fig:coolplot} contains all the correlations between the various parameters.~The true value of $A_2^{\gamma\gamma} = -0.008$ is shown by the black solid line while zero is given by the dashed line.~The colors indicate the density of pseudoexperiments which return a particular value of $A_2^{\gamma\gamma}$ with red being high density and blue being low.~We see clearly that the fit is sensitive to the sign of $A_2^{\gamma\gamma}$.

\begin{figure}[t]
\vspace{-0.75cm}
\includegraphics[width=.52\textwidth]{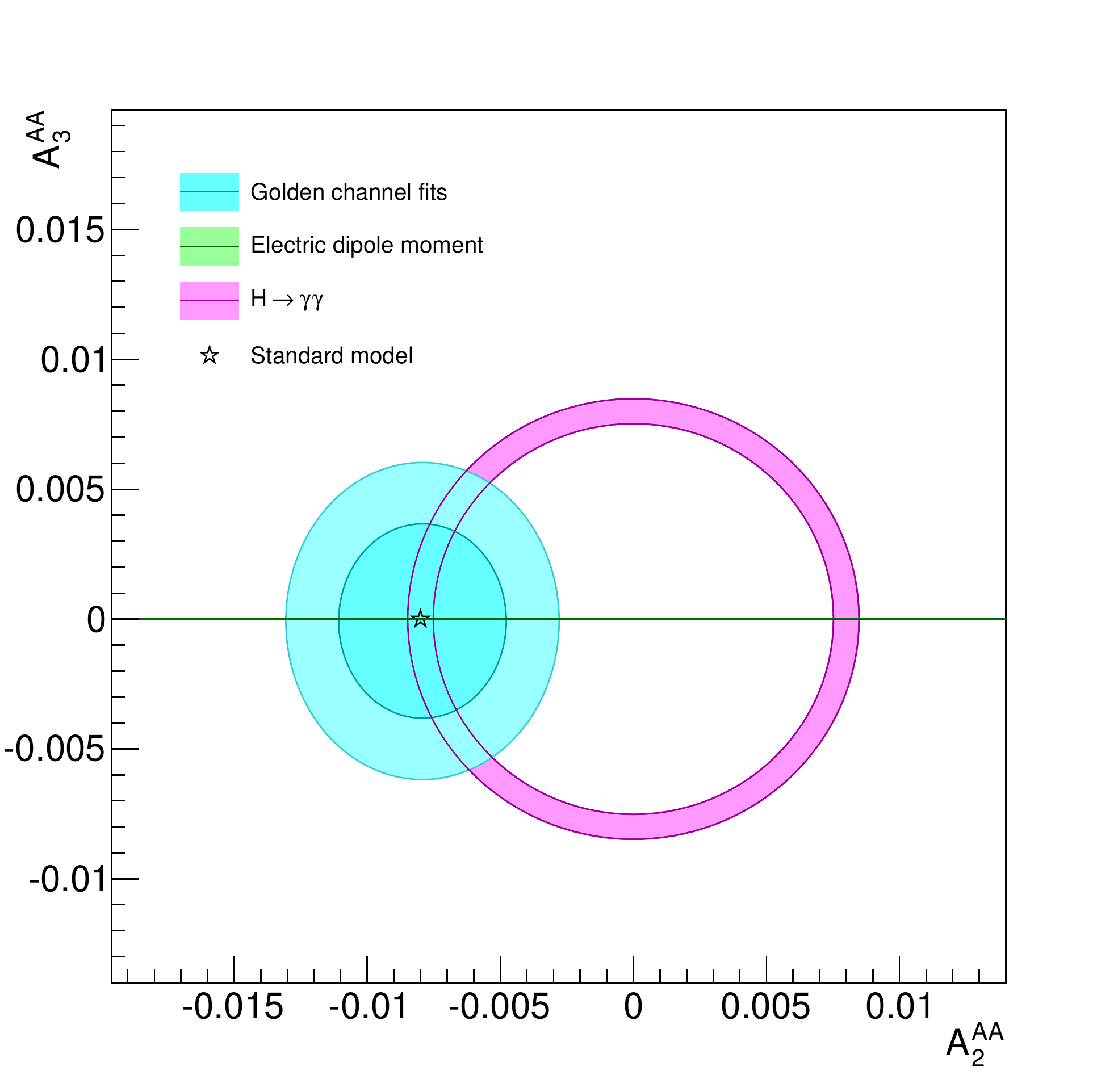}
\caption{The results of our parameter extraction for the true `SM-like' point represented by the star and defined in Eq.(\ref{eqn:A2AA}).~We combine the $2e2\mu$, $4e$, and $4\mu$ channels in the data set and our likelihood.~The fit is performed for $N_S = 12800$ events which roughly corresponds to the projected final LHC luminosity of $\sim 3000 fb^{-1}$ and assumes a uniform efficiency of $\sim 60\%$ for all three final states.~We show fit results for $A_2^{\gamma\gamma}$ vs $A_3^{\gamma\gamma}$ (floating all couplings).~The turquoise circles correspond to the $68\%$ and $95\%$ confidence interval obtained in the golden channel while the pink ring indicates the projected $1\sigma$ confidence interval which will be achieved in $h\rightarrow\gamma\gamma$~\cite{htoAA} for the same luminosity.~The thin green line shows the severe constraint coming from the electron EDM in a minimal model where the mass of the states which generate these operators is a TeV and that the Higgs couplings to first generation fermions are of order their SM value~\cite{McKeen:2012av,Baron:2013eja}.~This constraint can be completely relaxed in other models~\cite{McKeen:2012av}.}
\label{fig:moneyplot}
\end{figure}

A further demonstration of the ability of the golden channel to extract the CP properties of the Higgs couplings to photons is shown in Fig.~\ref{fig:moneyplot}.~We show the results for a large set of pseudoexperiments each containing $12800$ events.~This corresponds roughly to an integrated luminosity of $3000 fb^{-1}$ assuming a uniform efficiency of $60\%$ for all three final states.~We show fit results in the 2D plane for $A_2^{\gamma\gamma}$ vs $A_3^{\gamma\gamma}$ where the turquoise circles correspond to the $68\%$ and $95\%$ confidence intervals obtained in our fit.~The pink ring indicates the projected $1\sigma$ confidence interval which will be achieved in the $h\rightarrow\gamma\gamma$ decay channel~\cite{htoAA} for the same luminosity.~The pink ring makes it clear that the $h\to\gamma\gamma$ process is only sensitive to the combination $|A_2^{\gamma\gamma}|^2+|A_3^{\gamma\gamma}|^2$ and thus can not directly probe the CP nature of these couplings.~We also show in the thin green line the very strong, but highly model dependent, constraint coming from the electron EDM~\cite{McKeen:2012av,Baron:2013eja}.~For this constraint we have assumed the couplings of the Higgs to first generation fermions is of order their SM value and that the mass of the states which generate these operators is a $\sim$ TeV.~However, the green line makes it clear that even with these model dependent assumptions, EDM measurements can not establish the overall sign of the Higgs photon coupling.

We see very simply from Fig.~\ref{fig:moneyplot} that by the end of running, the LHC should be able to establish the CP nature of the Higgs to photon couplings in the golden channel and in particular be able to determine the overall sign of the $A_2^{\gamma\gamma}$ coupling.~As can also be seen from Fig.~\ref{fig:moneyplot} this is something that can not be unambiguously established using the $h\rightarrow\gamma\gamma$ channel and EDM measurements.~This makes the golden channel the unique method capable of determining these properties in the foreseeable future.

\subsection{Comments on Results and Approximations}
\label{subsec:approx}

Of course the results we have presented in this study are the ideal case.~We have used simply the LO fully differential cross section for $h\rightarrow 4\ell$ and performed fits to data generated from the analytic expression itself.~There are a number of additional effects we neglected including production, background, and NLO decay effects~\cite{Bredenstein:2006rh,Bredenstein:2006nk}.~However, all of these effects which we have neglected are sub dominant~\cite{Chen:2012jy,Chen:2013ejz,Chen:2014pia} and do not become important until we begin to reach the level of sensitivity needed to measure the SM prediction for the effective Higgs couplings.~Thus they do not qualitatively change the results presented here and in particular the conclusion that the LHC has excellent prospects of directly establishing the CP properties of the Higgs coupling to photons.~To make more precise statements a more detailed framework is needed which includes these various effects.~We believe much progress can be made on all of these aspects and that sensitivity closely approximating Figs.~\ref{fig:LuminosityPlots}-\ref{fig:moneyplot} can be achieved during LHC running.~We leave a study of all of these effects to ongoing work~\cite{followup} building on the framework introduced in~\cite{Chen:2012jy,Chen:2013ejz,Chen:2014pia}.
\\~\\
\section{Conclusions}
\label{sec:conclusion}

We have examined the expected sensitivity of the $h\to 4\ell$ golden channel to the loop induced couplings of the Higgs boson to $ZZ$, $Z\gamma$, and $\gamma\gamma$ gauge boson pairs for values approximating those predicted by the Standard Model.~We have demonstrated qualitatively that the golden channel has excellent prospects of directly establishing the CP nature of the Higgs couplings to photons, well before the end of LHC running, with less optimistic prospects for the $ZZ$ and $Z\gamma$ loop induced couplings.

We emphasize that in obtaining our results we have not attempted to optimize the analysis for sensitivity to these couplings.~As part of an ongoing investigation we examine whether by altering the cuts and reconstruction which are applied one can enhance the sensitivity even more than what has been found here.

Even without optimizing our analysis we find that with standard `CMS-like' cuts and reconstruction and with $\sim 100-400 fb^{-1}$ of luminosity the LHC will reach the levels necessary to begin probing the loop induced Standard Model effects which generate the Higgs coupling to photons.~This of course warrants further study, but indicates that the golden channel is capable of directly probing the CP properties of the Higgs couplings to photons, including the overall sign, by the end of LHC running.

This measurement can not be made in the $h\rightarrow \gamma\gamma$ channel or in other indirect approaches without making model dependent assumptions.~This makes the golden channel the unique method capable of determining these properties in the foreseeable future and we encourage experimentalists at the LHC to carry out this measurement.
\\
\\
\noindent
{\bf Acknowledgments:}~We thank Ian Low, Joe Lykken, and Maria Spiropulu for providing us with the resources necessary to complete this study and for valuable discussions.~We also thank Michalis Bachtis, Emanuele Di Marco, Adam Falkowski, Mateo Garcia, Fabio Maltoni, Pedro Schwaller, Daniel Stolarski, Nhan Tran, Roberto Vega, Mark Wise, and Si Xie for helpful discussions.~R.V.M.~is supported by the Fermilab Graduate Student Fellowship in Theoretical Physics and the ERC Advanced Grant Higg@LHC.~Fermilab is operated by Fermi Research Alliance, LLC, under Contract No.~DE-AC02-07CH11359 with the United States Department of Energy.~Y.C.~is supported by the Weston Havens Foundation and DOE grant No.~DE-FG02-92-ER-40701.~This work is also sponsored in part by the DOE grant No.~DE-FG02-91ER40684.~This work used the Extreme Science and Engineering Discovery Environment (XSEDE), which is supported by National Science Foundation grant number OCI-1053575.


\bibliographystyle{apsrev}
\bibliography{GoldenChannelBibNew}

\end{document}